\documentstyle[12pt,aaspp4,flushrt]{article}
\def\etal{{et al.}}
\def\name{SDSS 1533--00}
\begin{document}

\title{The Discovery of a High-redshift Quasar without Emission Lines from Sloan
Digital Sky Survey Commissioning Data\footnote{Based on observations obtained with the
Sloan Digital Sky Survey and the Apache Point Observatory
3.5-meter telescope, which are owned and operated by the Astrophysical
Research Consortium, the Very Large Array, the Keck Telescope, and the
Steward Observatory 2.3m Telescope.}}
\author{
Xiaohui Fan\altaffilmark{\ref{Princeton}},
Michael A. Strauss\altaffilmark{\ref{Princeton}},
James E. Gunn\altaffilmark{\ref{Princeton}}, 
Robert H. Lupton\altaffilmark{\ref{Princeton}}, 
C. L. Carilli\altaffilmark{\ref{NRAO}},
M. P.  Rupen\altaffilmark{\ref{NRAO}},
Gary D. Schmidt\altaffilmark{\ref{UA}},
Leonidas A. Moustakas\altaffilmark{\ref{UCB},\ref{Oxford}},
Marc Davis\altaffilmark{\ref{UCB}},
James Annis\altaffilmark{\ref{Fermilab}},
Neta A. Bahcall\altaffilmark{\ref{Princeton}},
J. Brinkmann\altaffilmark{\ref{APO}},
Robert J. Brunner\altaffilmark{\ref{JHU},\ref{Caltech}},
Istv\'an Csabai\altaffilmark{\ref{JHU},\ref{Eotvos}},
Mamoru Doi\altaffilmark{\ref{UTokyo}}
Masataka Fukugita\altaffilmark{\ref{CosmicRay},\ref{IAS}},
Timothy M. Heckman\altaffilmark{\ref{JHU}},
G. S. Hennessy\altaffilmark{\ref{USNO}},
Robert B. Hindsley\altaffilmark{\ref{USNO}},
\v{Z}eljko Ivezi\'{c}\altaffilmark{\ref{Princeton}},
G. R. Knapp\altaffilmark{\ref{Princeton}},
D. Q. Lamb\altaffilmark{\ref{Chicago}},
Jeffrey A. Munn\altaffilmark{\ref{Flagstaff}},
A. George Pauls\altaffilmark{\ref{Princeton}},
Jeffrey R. Pier\altaffilmark{\ref{Flagstaff}},
Constance M. Rockosi\altaffilmark{\ref{Chicago}},
Donald P. Schneider\altaffilmark{\ref{PennState}},
Alexander S. Szalay\altaffilmark{\ref{JHU}},
Douglas L. Tucker\altaffilmark{\ref{Fermilab}},
Donald G. York\altaffilmark{\ref{Chicago}}
}

\newcounter{address}
\setcounter{address}{2}
\altaffiltext{\theaddress}{Princeton University Observatory, Princeton, NJ 08544
\label{Princeton}}
\addtocounter{address}{1}
\altaffiltext{\theaddress}{NRAO, P.O. Box O, Socorro, NM 87801
\label{NRAO}}
\addtocounter{address}{1}
\altaffiltext{\theaddress}{Steward Observatory, University of Arizona, 933 North Cherry Avenue, Tucson, AZ 85721
\label{UA}}
\addtocounter{address}{1}
\altaffiltext{\theaddress}{Department of Astronomy, University of California, Berkeley, CA 94720-3411
\label{UCB}}
\addtocounter{address}{1}
\altaffiltext{\theaddress}{Dept.~of Astrophysics, Oxford University, Oxford OX1 3RH, UK
\label{Oxford}}
\addtocounter{address}{1}
\altaffiltext{\theaddress}{Fermi National Accelerator Laboratory, P.O.
 Box 500,
Batavia, IL 60510
\label{Fermilab}}
\addtocounter{address}{1}
\altaffiltext{\theaddress}{Apache Point Observatory, P.O. Box 59,
Sunspot, NM 88349-0059
\label{APO}}
\addtocounter{address}{1}
\altaffiltext{\theaddress}{
Department of Physics and Astronomy, The Johns Hopkins University,
   3701 San Martin Drive, Baltimore, MD 21218, USA
\label{JHU}}
\addtocounter{address}{1}
\altaffiltext{\theaddress}{
Department of Astronomy, California Institute of Technology,
Pasadena, CA 91125
\label{Caltech}
}
\addtocounter{address}{1}
\altaffiltext{\theaddress}{Department of Physics of Complex Systems,
E\"otv\"os University,
   P\'azm\'any P\'eter s\'et\'any 1/A, Budapest, H-1117, Hungary
\label{Eotvos}
}
\addtocounter{address}{1}
\altaffiltext{\theaddress}{Department of Astronomy and Research Center for the Early Universe, School of Science, University of Tokyo, Hongo, Bunkyo,
Tokyo, 113-0033 Japan
\label{UTokyo}}
\addtocounter{address}{1}
\altaffiltext{\theaddress}{Institute for Cosmic Ray Research, University of
Tokyo, Midori, Tanashi, Tokyo 188-8502, Japan
\label{CosmicRay}}
\addtocounter{address}{1}
\altaffiltext{\theaddress}{Institute for Advanced Study, Olden Lane,
Princeton, NJ 08540
\label{IAS}}
\addtocounter{address}{1}
\altaffiltext{\theaddress}{U.S. Naval Observatory,
3450 Massachusetts Ave., NW,
Washington, DC  20392-5420
\label{USNO}}
\addtocounter{address}{1}
\altaffiltext{\theaddress}{University of Chicago, Astronomy \& Astrophysics
Center, 5640 S. Ellis Ave., Chicago, IL 60637
\label{Chicago}}
\addtocounter{address}{1}
\altaffiltext{\theaddress}{U.S. Naval Observatory, Flagstaff Station,
P.O. Box 1149,
Flagstaff, AZ  86002-1149
\label{Flagstaff}}
\addtocounter{address}{1}
\altaffiltext{\theaddress}{Department of Astronomy and Astrophysics,
The Pennsylvania State University,
University Park, PA 16802
\label{PennState}}

\begin{abstract}
We report observations of a luminous unresolved object at redshift $z
= 4.62$, with a featureless optical spectrum redward of the Lyman
$\alpha$ forest region, discovered from Sloan Digital Sky Survey
(SDSS) commissioning data.  The redshift is determined by the onset of
the Lyman-$\alpha$ forest at $\lambda \sim 6800$ \AA, and a Lyman
Limit System at $\lambda = 5120$ \AA.  A strong Ly$\alpha$ absorption
system with weak metal absorption lines at $z=4.58$ is also identified
in the spectrum.  The object has a continuum absolute magnitude of
$-26.6$ at 1450\AA\ in the rest-frame ($h_{0}=0.5$, $q_0=0.5$), 
and therefore cannot be an ordinary galaxy.  
It shows no radio emission (the 3$\,\sigma$ upper limit of its flux at 6
cm is 60$\mu$Jy), indicating an radio-to-optical flux ratio at least
as small as that of the radio-weakest known BL Lacs. 
It is also not linearly polarized in the observed $I$ band to a
3$\,\sigma$ upper limit of 4\%.  
Therefore, it is either the most distant BL Lac object known to date,
with very weak radio emission, 
or a new type of unbeamed quasar, 
whose broad emission-line region is very weak or absent.
\end{abstract}

\section{Introduction}
The Sloan Digital Sky Survey (SDSS; 
\cite{GW95}; \cite{York99}\footnote{see also
\texttt{http://www.astro.princeton.edu/PBOOK/welcome.htm}}) is using a
dedicated 2.5m telescope at Apache Point Observatory, New Mexico and a
wide-field camera with 54 CCDs (Gunn \etal\ 1998), to
obtain CCD images in five broad optical bands ($u'$, $g'$, $r'$, $i'$,
$z'$, centered at 3540\AA, 4770\AA, 6230\AA, 7630\AA\ and 9130\AA,
\cite{Fukugita96}) over 10,000 deg$^2$ of the high Galactic latitude
sky centered approximately on the North Galactic Pole.  The multicolor
data allow objects of spectral energy distribution distinct from that
of stars to be easily selected as outliers 
from the stellar locus in color-color space.  In particular, the SDSS
has proven to be a 
successful source of discoveries of high-redshift quasars, selected by
their distinctive colors in the redder filters.  \cite{Fan99} has
shown that for redshifts greater than 2, the colors of quasars in the
SDSS filter system are a strong function of redshift, as first the
Lyman-$\alpha$ forest, and then the Lyman Limit Systems (LLSs), move
through the filter system.  For redshifts appreciably above 4, quasars
are essentially invisible in the SDSS $u'$ and $g'$ filters, and thus
must be selected from their $r'-i'$ and $i'-z'$ colors.  In the
redshift range from 3.6 to 5, the distinction between the stellar and quasar
loci is quite sharp, and follow-up spectroscopy of early SDSS imaging
commissioning data has revealed 40 previously unknown
quasars in this redshift range (Fan \etal\ 1999a, b, \cite{HET}).

Spectroscopic follow-up of outliers from the stellar locus has allowed
the discovery of objects other than ordinary quasars.  In this paper,
we present the discovery of a luminous point source at $z = 4.62$, which in
contrast to ordinary quasars, shows no emission lines. 

\section{Observations}
The object SDSSp J153259.96--003944.1
(which we refer to hereafter as \name\ for brevity) was selected
from SDSS Photometric Run 77 (observed on 27 June 1998) 
as a high-redshift quasar candidate.
The photometric observations and target selection of this run
are described in detail in Fan et al.~(1999b).
This region of the sky was re-imaged in Run 752 (21 Mar 1999).
Photometric calibration for these data was
provided by an auxiliary $24''$ telescope (now decommissioned) at the
same site.  The survey data 
processing software carries out astrometric and photometric
calibrations, and finds and measures properties of all detected objects in the
data (\cite{ASTROM}\footnote{see also {\tt
http://www.astro.princeton.edu/PBOOK/astrom/astrom.htm}},
\cite{Lupton99b}\footnote{see also {\tt
http://www.astro.princeton.edu/PBOOK/datasys/datasys.htm}}).

Table 1 gives the results of the astrometry
and photometry in these two observations of \name.  As this table
shows, the photometry is consistent between two dates separated by 9
months in the bands in which it was detected at high signal-to-noise
ratio.  The absolute calibration of the photometry is 
uncertain at the 5\% level, as the primary standard star network had
not been completely established when these data were taken; for this
reason, we indicate our photometry with asterisks rather than the
primes of the final system, although we continue to refer to the
filters themselves with the prime notation.  
A finding chart for \name\ in the $i'$ band is given in Fan et
al.~(1999b). 
The object was undetected in $u'$ and $g'$, and has colors
$r^*-i^* \sim 1.4$, $i^*-z^* \sim 0.2$, typical for a quasar
at $z \sim 4.7$ (see Figure 1 of Fan et al.~1999b).

  We obtained optical spectra of \name\ on 13 March 1999 UT using the
Double Imaging Spectrograph on the Apache Point 3.5m telescope, with
the same instrumental configuration used by Fan \etal\ (1999a,b).  
The resolution is 13\AA, and the spectral coverage is 4000--10,000\AA.
Observations of the spectrophotometric standard BD$\,$+33$\,^\circ2642$
(\cite{Oke90}) provided flux calibration and allowed removal of
the atmospheric absorption bands.  
The seeing was about $1.4''$ on this photometric night, 
and the observations were carried out at low airmass. 
The resulting spectrum, a co-addition of
three 30 minute exposures, 
is shown in Figure~1; this spectrum also
appears in Fan \etal\ (1999b).

The optical spectrum of this object is very unusual; 
it has no emission lines,  and possesses two distinct breaks. 
The first break is at $\sim 6800$\AA.
Redward of this break, the spectrum shows only a smooth 
power-law-like continuum,
with no obvious emission line or strong absorption features.
Blueward of this break, it shows numerous strong absorption
lines typical of the Lyman $\alpha$ forest region in high-redshift quasars. 
The second break is at 5120 \AA, blueward of which there is no
detectable flux.
If these two breaks are the onset of Lyman $\alpha$ and a strong
LLS respectively,  they give a consistent redshift of 4.62.
Furthermore, a damped Ly$\alpha$ system candidate at $z=4.58$ is 
identified by the strong Ly$\alpha$ and Ly$\beta$ and weak
Ly$\gamma$ absorption lines in the spectrum, as indicated in the figure.
The redshift determination of \name\ 
and the nature of the Ly$\alpha$ system
is further discussed in \S 3.

Given the unusual nature of the spectrum, we immediately 
obtained additional observations.  We obtained deep VLA imaging of the
field of \name\ on 3 April 1999 in the D array. These observations
used two circular polarizations and 
two 50 MHz IF's at 4860 MHz separated by 50 MHz, for a total effective
bandwidth of 100 MHz in two orthogonal polarizations. The total
integration time was 72 minutes.  The image was CLEANed with a
Gaussian restoring beam of $20'' \times 14''$  at PA = $34^\circ$. The
resulting map has a 1$\,\sigma$ noise 
level of 20$\mu$Jy per beam; no source is detected above 3$\,\sigma$ at the
position corresponding to our object.
Moreover, this object is also not detected on the ROSAT full-sky pixel images (
W. Voges, private communication, \cite{ROSAT}),
implying a 3$\,\sigma$ upper limit of X-ray flux of  $\rm 3 \times 10^{-13}\
erg\ cm^{-2}s^{-1}$ in the 0.1 -- 2.4 keV band. 

We obtained optical polarimetric observations of \name\
using the Steward Observatory 2.3m Bok telescope and
imaging/spectropolarimeter 
(\cite{Schmidt92}) on the night of 18 May 1999.  The
instrument images both polarized beams simultaneously; chopping 
was done via rotation of a half-waveplate every 60 sec.  Our total
integration was 3840 sec.  The Hoya R72 filter 
used cuts off at the blue end at $\approx 7200$\AA; the red end 
of the passband was 
defined by the CCD sensitivity function, which declines very quickly beyond
$\approx 8000$\AA.  The seeing on this night varied between 1.0 and 1.5
arcsec.
Our final result
is a linear polarization of $P = 0.7\% \pm 1.3\%$, after correction for the
uncertainty bias in polarization measurements; the 
uncertainty was estimated from the variance in results from 
four independent measurements.  We conclude that the object is unpolarized 
to a 3$\,\sigma$ upper limit of 4\%.

Finally, we obtained higher signal-to-noise ratio (S/N) spectroscopy of
\name\ with the Keck II telescope.  The object was observed with the Low
Resolution Imaging Spectrograph (LRIS, \cite{LRIS}) on the night of
13 May 1999.  The resolution was roughly 7\AA, and the spectra covered
the range from 5100-9000\AA.  We obtained a 2700 second exposure,
under photometric skies and good seeing. 
The spectrum is shown in Figure 2.
The Keck spectrum is consistent with the APO spectrum, with all the
spectral features in this wavelength range reproduced.

\section{Discussion}

\subsection{Redshift Determination}

The interpretation of object \name\ as a high-redshift object is unambiguous.
In addition to the presence of two breaks in the spectrum representing
the onset of the Lyman $\alpha$ forest and the LLS absorption, the
properties of the absorption lines are consistent with our interpretation.
We calculate the flux decrement parameters $D_{A}$ and $D_{B}$ 
(\cite{OK82}), the average flux depression due to absorption lines
in the rest-frame wavelength ranges 1050-1170 \AA, and 920--1015\AA.
We determine the continuum level by fitting a power law for 
$\lambda > 7200$\AA. Assuming $z=4.62$,
we  get $D_{A}=0.63$ and $D_{B}=0.69$, very typical values for objects at similar redshift
(e.g., \cite{SSG}, \cite{APMLLS}, \cite{Kennefick95}).

In Figure 2, we examine the property of the Ly$\alpha$ absorption 
system at $z=4.58$ in detail, using the Keck spectrum. 
A Voigt profile with $N_{HI}= 2.5\times 10^{20}$ cm$^{-2}$
is overplotted on the observed Ly$\alpha$ absorption line.
Although the shape of the red wing of the absorption line matches 
the damped wing of the Voigt profile,
the core of the line is wider, and it rises much more steeply
than does the Voigt profile.
Furthermore, the core of the Ly$\beta$ absorption line does not
reach zero flux.
These facts indicate the absorption system has more complicated
velocity structure, rather than being a single damped Ly$\alpha$ absorber. 
Weak metal absorption lines of Si$\,$II$\lambda\, 1260.4$ and 
C$\,$II$\lambda\, 1334.5$ are tentatively detected in the spectrum.
The Si$\,$IV and C$\,$IV absorption lines would be in the more noisy
part of the spectrum and are not clearly detected.
Future observations with higher resolution and S/N are needed
to fully understand this absorption system and whether it is physically
related to \name\ itself.
We also note a broad ``emission'' feature at $\sim 5800$ \AA\ in
Figures 1 and 2. 
This wavelength coincides with  that of O$\,$VI$\lambda\, 1034$ at $z
\sim 4.6$, but this may just be a void in the Ly$\alpha$ forest. 

Because of the absence of emission lines, determination of the exact 
redshift of \name\ is not straightforward.
The existence of the Ly$\alpha$ absorption system 
at $z_{\rm abs}=4.58$
suggests that the intrinsic redshift $z \gtrsim 4.58$.
The presence of the LLS at 5120\AA\ indicates $z \gtrsim 4.62$.
The number density of LLSs increases with redshift.
For quasars at this redshift, the redshift of the LLS is typically within
0.1 of the emission line redshift (\cite{SSG}, \cite{APMLLS}, Fan \etal\ 1999a,b).
The high S/N Keck spectrum shows no obvious absorption due to
Lyman $\alpha$ forest lines beyond 6900 \AA, indicating that 
the redshift is smaller than 4.67.
We therefore assign a redshift of $4.62 \pm 0.04$. 
A more accurate redshift can be determined if emission lines are
detected at infrared wavelengths.

\subsection{Could This be a BL Lac Object?}

The absolute AB magnitude of \name\ at 1450\AA\ (rest-frame) is $\rm
M_{1450}=-26.59$ (for $q_{0}=0.5$, $h_{0}=0.5$, after correcting for
Galactic extinction using the reddening map of Schlegel, Finkbeiner,
\& Davis 1998). 
Thus, although the S/N of our optical spectrum is probably not high
enough to rule out the presence of stellar absorption lines, this
object is much too luminous to be an ordinary galaxy.  It shows no
broad emission lines in its optical spectrum, and therefore is not a
typical quasar.  A Ly$\alpha$ emission with rest-frame equivalent
width of 5\AA\ would be easily detected in our spectrum.  For comparison, the
typical quasar at $z \sim 4.6$ has a Ly$\alpha$ rest-frame equivalent
width of $\sim 60$\AA\ (e.g., Fan \etal\ 1999a,b); the lowest
equivalent width of Ly$\alpha$ in a high-redshift quasar of which we
are aware is PSS 1435+0357 at $z = 4.35$ (\cite{Kennefick95}), which has a 10\AA\
line in the rest frame.  It is also unlikely that
the emission lines are severely affected by the $z=4.58$ absorption
system.  Even if the Ly$\alpha$ emission were partly absorbed, the
absorption lines associated with Si$\,$II and C$\,$II are weak,
arguing that the effect of the absorption due to C$\,$IV and Si$\,$IV
on any putative emission lines there would be weak. 

There is no
indication that \name\ is extended beyond the point spread function of
the $1.3''$ FWHM SDSS image.  Thus this object cannot be a
normal galaxy amplified by gravitational lensing. 
Therefore \name\  is either 
an AGN whose emission lines have been swamped by
a relativistically beamed continuum (i.e., a BL Lac or Blazar), or there is
simply no broad emission line region at all.

A BL Lac object shows no or very weak emission lines (\cite{UP95}). 
It is also characterized by strong radio and X-ray emission,
optical variability, and strong and variable optical polarization 
due to the synchrotron radiation from the relativistic jet. 
No radio-quiet BL Lac has ever been found
(\cite{Stocke90}, \cite{JGF93}), in the sense that every BL Lac for
which mJy level radio data exist has been detected.
The highest redshift BL Lac in the literature is at $z<2$
(\cite{LM99}).

However, other than the absence of optical emission lines, \name\
exhibits none of the properties that define a BL Lac object, 
although the current observations cannot definitively rule out that it
is a BL Lac: (1) it is radio-quiet
at low flux level.  The 3$\,\sigma$ limit at 6 cm is 60 $\mu$Jy.
The $i^*$ magnitude of $19.7$ corresponds to a flux of 48  $\mu$Jy
at $\rm \lambda_{eff} = 7600$\AA.
This implies a radio-optical spectral index in the observed frame 
$\rm \alpha_{ro} < 0.02$ at 3\,$\sigma$. 
Classical radio-selected BL Lacs have $\rm
\alpha_{ro} > 0.3$ (\cite{Stocke90}), and X-ray-selected BL Lacs can
have $\rm \alpha_{ro}$ as small as $\sim 0.1$ (\cite{LM99}). 
This indicates an radio-to-optical flux ratio of \name\ 
smaller than that of BL Lacs with the weakest known radio emission.
However, the  radio K-correction of BL Lacs is completely unknown at
this redshift.  
Therefore, it is still possible that the radio-to-optical flux ratio of \name\
is comparable to that of a low-redshift BL Lac with very weak radio emission.
Yet deeper radio imaging is needed to put further constraints on its
radio properties. 
(2) It is not optically polarized (the
3$\,\sigma$ limit on its linear polarization is 4\%).  BL Lacs have
maximum linear polarization of order 10\%, and the duty cycle for
polarization (i.e., the fraction of the time that any given object has
a polarization greater than 4\%) is 40--60\% (\cite{KS90},
\cite{JSE94}), so it is possible that we have observed it on its
off cycle. 
(3) The photometry of \name\ at two epochs separated by 9 months (1.5 months
in the rest-frame of the object) is consistent to within 0.05 mag
in $r'$, $i'$ and $z'$ bands.
Further long-term monitoring of \name\ is needed to put a
stronger constraint on its polarimetric and photometric variability.
(4) It is not detected in ROSAT X-ray full-sky images.  The
optical-X-ray index in the observed frame is $\rm \alpha_{ox} > 1.10$ 
at $3\,\sigma$ (assuimg an X-ray spectral index $\rm \alpha_x = 1$). 
This lower limit is in the middle of the distribution of $\rm \alpha_{ox}$
in the samples of Stocke et al.~(1990), and 
values of $\rm \alpha_{ox}$ as large as $\sim 2$ are seen in
radio-selected BL Lac samples (e.g. \cite{LM99}). Therefore, the X-ray
upper limit is not a strong constraint. 

Thus, three interpretations present themselves.  First, \name\ could
be the most distant BL Lac object known to date.  If this is true, it is
unusually radio weak, but is likely to show a faint radio counterpart
with further observing.  It should also show signatures of
relativistic beaming such as strong photometric and polarimetric variability.
A population of high-redshift BL Lacs has in fact long been speculated
by Stocke \& Perrenod (1981) and Stocke (1989). 
Second, it is conceivable that dust
simply extincts the broad-line region of a normal quasar in this object, 
but the continuum slope of this
object, as measured from the $i^*-z^*$ color, is comparable to other
quasars at a similar redshift (Fan \etal\ 1999a,b), and thus shows no
obvious sign of reddening.  Moreover, resonant scattering of the Ly$\alpha$
line to reduce the requirement on the dust column density would not
effectively extinct the broad wings of the line. 
Finally, \name\ could be a new kind
of quasar with no or very weak broad emission line region.
Such objects are clearly very rare; broad-band color selection of
high-redshift objects is not very sensitive to the emission line
strength (Fan 1999), yet \name\ is the only quasar without emission lines
among the 80 or so $z>4$ quasars found in multicolor surveys (Schneider 1999,
Fan  \etal\ 1999a,b).
Further observations, including spectroscopy in the infrared to 
determine the UV/optical/IR spectral energy distribution of the object
and to look for other emission lines, HST imaging to look for
extended emission from the object, even deeper radio imaging, 
deep X-ray observations with XMM or AXAF, 
and long term optical monitoring for photometric and polarimetric variability
will help us to understand the physical nature of \name.

The Sloan Digital Sky Survey (SDSS) is a joint project of the
University of Chicago, Fermilab, the Institute for Advanced Study, the
Japan Participation Group, The Johns Hopkins University, the
Max-Planck-Institute for Astronomy, Princeton University, the United
States Naval Observatory, and the University of Washington.  Apache
Point Observatory, site of the SDSS, is operated by the Astrophysical
Research Consortium.  Funding for the project has been provided by the
Alfred P. Sloan Foundation, the SDSS member institutions, the National
Aeronautics and Space Administration, the National Science Foundation,
the U.S. Department of Energy, and the Ministry of Education of Japan.
The SDSS Web site is {\tt http://www.sdss.org/}.  XF and MAS acknowledge
additional support from Research Corporation, 
NSF grant AST96-16901, the Princeton University Research Board, and 
an Advisory Council Scholarship.  We thank John Bahcall, Bruce Draine,
Gordon Richards, Todd Tripp, Wolfgang Voges, and the referee, John
Stocke, for useful discussions and comments, and Russett McMillan for her
usual expert help at the 3.5m telescope.

\newpage

\begin{footnotesize}
\begin{deluxetable}{lcccccc}
\tablenum{1}
\tablecolumns{7}
\tablecaption{Optical Positions and SDSS Photometry of SDSS 1533-00}
\tablehead
{
Position (J2000)& $u^*$ & $g^*$ & $r^*$ & $i^*$ & $z^*$ &Date
}
\startdata
15:32:59.96 --00:39:44.1 & 24.60 $\pm$ 0.46  & 23.18 $\pm$ 0.27 & 21.19 $\pm$ 0.06 & 19.73 $\pm$ 0.03  & 19.54 $\pm$ 0.11 & 27 June 1998\\
15:32:59.93 --00:39:43.9 & 23.70 $\pm$ 0.39  & 23.81 $\pm$ 0.30 & 21.15 $\pm$ 0.05 & 19.75 $\pm$ 0.03 & 19.57 $\pm$ 0.07 & 21 Mar 1999 \\
\enddata

\tablenotetext{}{Photometry is
reported in terms of {\em asinh magnitudes} on the AB system. 
The asinh magnitude system is defined in Lupton, Gunn \& Szalay (1999); 
it becomes a linear scale in flux when the absolute value of the
signal-to-noise ratio is less than about 5. In this
system, zero flux corresponds to 24.24, 24.91, 24.53, 23.89, and
22.47, in $u^*$, $g^*$, $r^*$, $i^*$, and $z^*$, respectively; larger
magnitudes refer to negative flux values.}
\end{deluxetable}
\end{footnotesize}

\begin{figure}
\vspace{-7cm}

\epsfysize=600pt \epsfbox{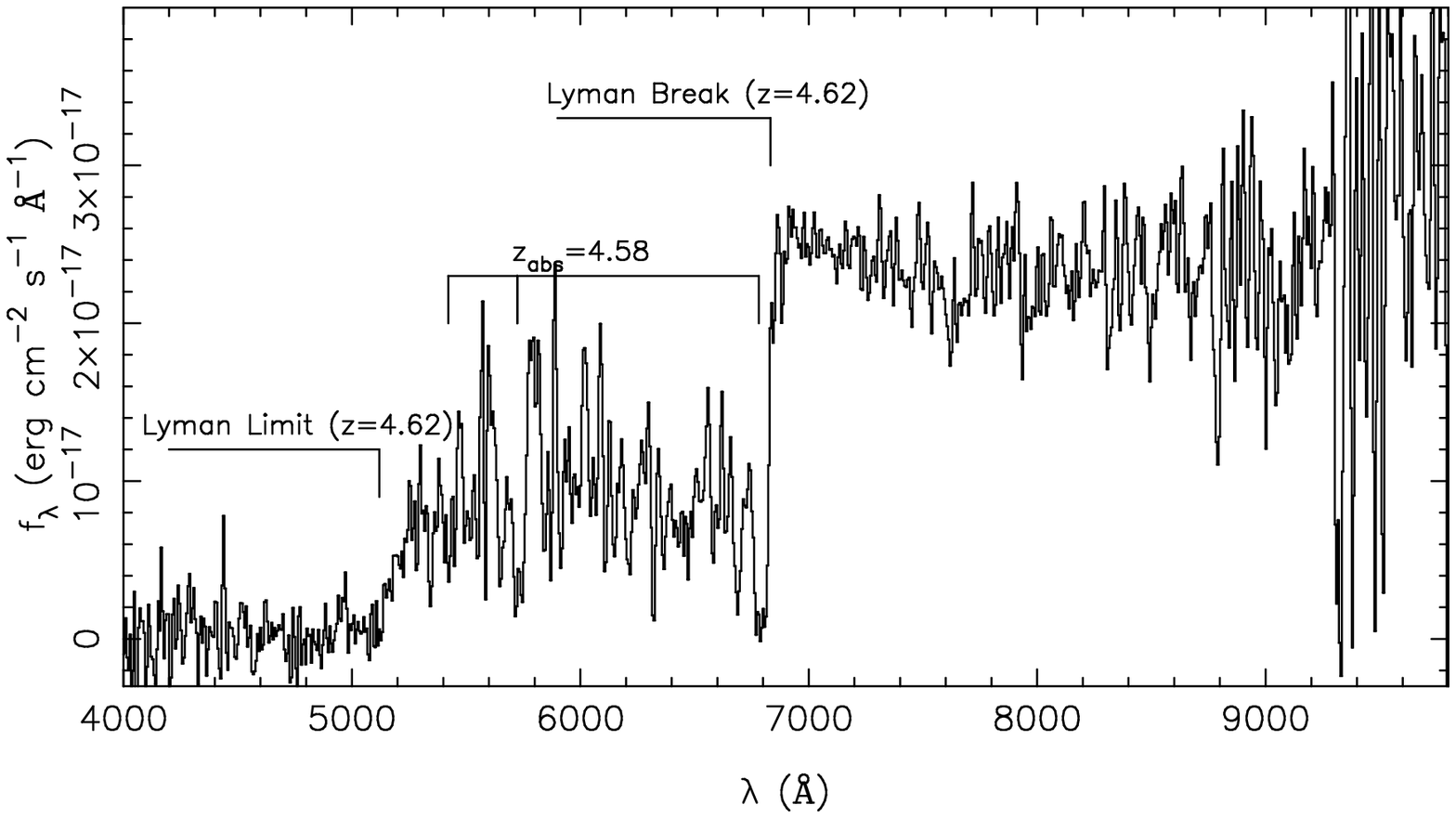}

\vspace{1cm}
Figure 1. Optical spectrum of \name\ obtained by the DIS spectrograph
on the ARC 3.5m telescope. 
The total exposure time is 5400 sec. The spectral resolution is
about 12 \AA\ in the blue and 14 \AA\ in the red. Each pixel
represents 6.2 \AA. The position of the Lyman break, Lyman limit, and
the Ly$\alpha$, $\beta$, and $\gamma$ lines of an absorption system at
$z = 4.58$ are indicated. 
\end{figure}

\begin{figure}
\vspace{-7cm}

\epsfysize=600pt \epsfbox{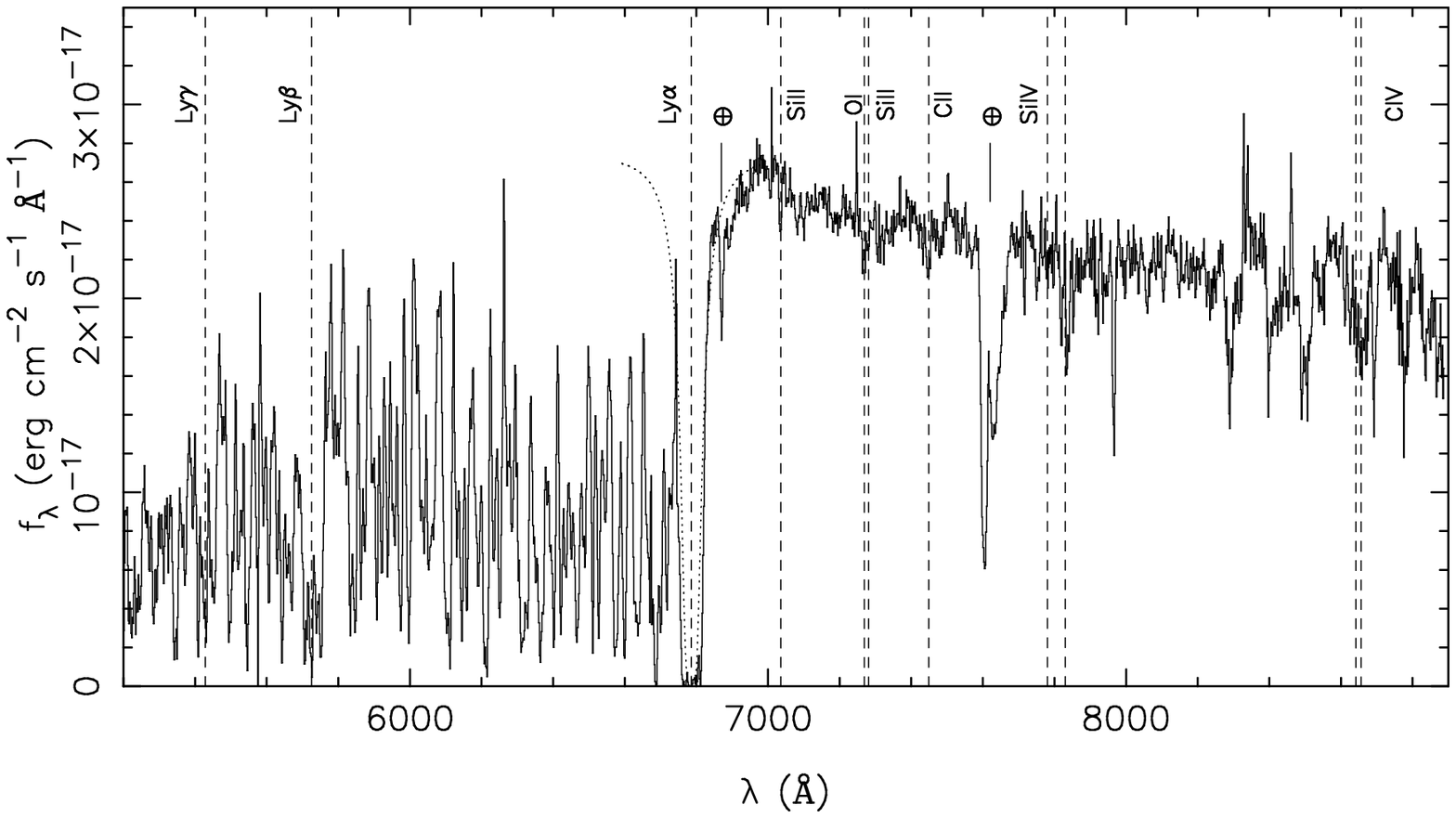}
\vspace{1cm} Figure 2. Optical spectrum of \name\ obtained by the LRIS
spectrograph on Keck II.  The total exposure time is 2700 sec.  The
spectral resolution is about 7\AA, and the dispersion is 1.9\AA\ per pixel.
No atmospheric absorption bands were removed from the spectrum.
Expected centroids of Ly$\alpha$, Ly$\beta$, Ly$\gamma$ and metal
absorption lines from the absorption system at $z=4.58$ are indicated.
The wing of the Ly$\alpha$ line is matched by a Voigt profile with
$N_{HI} = 2.5\times 10^{20}$ cm$^{-2}$.
\end{figure}
\end{document}